\documentclass[a4paper,11pt]{article}
\usepackage{aaskaiid}
\usepackage{orcidlink}
\usepackage{ulem}
\usepackage{CJK}

\usepackage{wrapfig}
\newcommand*{\rev}[1]{\textcolor{black}{#1}}
\newcommand*{\revtwo}[1]{\textcolor{black}{#1}}

\begin{document}
\begin{CJK*}{UTF8}{ipxm}
\title{Unravelling Turbulence and Magnetic Fields in Galaxy Clusters with SKA and XRISM}
\ShortTitle{ICM Magnetic Fields and Turbulence with SKA and XRISM}

\author[1]{Kohei Kurahara  (藏原~昂平)\orcidlink{0000-0003-2955-1239}}
\ShortName{Kurahara et al.} 
\author[2]{Takuya Akahori (赤堀~卓也) \orcidlink{0000-0001-9399-5331}}
\author[3]{Amit Seta \orcidlink{0000-0001-9708-0286}}
\author[4]{Kosuke Nishiwaki (西脇~公祐) \orcidlink{0000-0003-2370-0475}}
\author[1,5]{Kazuhiro Nakazawa (中澤~知洋) \orcidlink{0000-0003-2930-350X}}
\author[5]{Yuki Omiya (大宮~悠希) \orcidlink{0009-0009-9196-4174}}
\author[5]{Daisuke Ito (伊藤~大将) \orcidlink{0009-0000-4742-5098}}
\author[5]{Kosei Sakai (坂井~晃生)}

\affiliation[1]{Kobayashi-Maskawa Institute for the Origin of Particles and the Universe (KMI), Nagoya University, Furo-cho, Chikusa-ku, Nagoya, Aichi 464-8601, Japan}
\emailAdd{kurahara.kohei.i7@f.mail.nagoya-u.ac.jp}
\affiliation[2]{Mizusawa VLBI Observatory, National Astronomical Observatory of Japan, 2-21-1 Osawa, Mitaka, Tokyo 181-8588, Japan}
\affiliation[3]{Research School of Astronomy and Astrophysics, Australian National University, Canberra, ACT 2611, Australia}
\affiliation[4]{Istituto di Radioastronomia, INAF, Via Gobetti 101, 40121 Bologna, Italy}
\affiliation[5]{Department of Physics, Nagoya University, Furo-cho, Chikusa-ku, Nagoya, Aichi 464-8601, Japan}

\abstract{
This chapter proposes a research framework to quantitatively investigate non-thermal components in the Intracluster Medium (ICM) of galaxy clusters, which are critical ingredients for governing energy transport, structure formation, and particle acceleration. Turbulence, primarily driven by cluster mergers, is the leading mechanism for re-accelerating cosmic ray electrons (forming radio halos) and amplifying magnetic fields (via the turbulent dynamo).

Observational understanding of both the turbulence and magnetic fields is rapidly evolving: the high-resolution X-ray spectrometer \textbf{XRISM} is directly measuring the velocity properties of the thermal ICM, providing insights into the kinetic energy of turbulence. Concurrently, high-sensitivity low-frequency radio observations, including \textbf{SKA} pathfinders, are mapping non-thermal components and magnetic structures through diffuse synchrotron emission and high-density Faraday Rotation Measure (RM) grids.

The synergy between XRISM and SKA offers a decisive paradigm shift. XRISM's velocity maps, with its high energy resolution (<7 eV FWHM), combined with SKA-Mid's capability to deliver high-resolution RM grids ($\sim 100$--$200~\rm deg^{-2}$) and high-dynamic-range imaging, will allow for the first direct, multi-wavelength comparison of the turbulent energy properties (from X-ray) and the magnetic field properties (from radio). This joint analysis will validate Magnetohydrodynamic (MHD) simulation predictions, clarify the process of turbulent energy cascade and decay, and ultimately lead to a comprehensive understanding of the co-evolution of turbulence, magnetic fields, and cosmic rays in the largest laboratories of the Universe.
}


\maketitle

\section{Introduction}
Turbulence and magnetic fields are ubiquitous throughout the universe, playing fundamental roles across a broad range of spatial and temporal scales, from star formation and galaxy evolution to the emergence of large-scale cosmic structures \citep{2021amff.book.....S}. Elucidating their origin and evolution is essential for understanding the physical processes governing energy transport and structure formation in the universe. Among such cosmic systems, galaxy clusters constitute the largest self-gravitating structures in the universe and represent the final stage of hierarchical structure formation. A typical cluster comprises hundreds to thousands of galaxies, with the majority of its mass residing in dark matter and in the hot intracluster medium (ICM), whose temperature ranges from several tens to hundreds of millions of kelvins \citep[e.g.,][for a review]{2019SSRv..215...16V}. The ICM consists predominantly of high-temperature plasma and is primarily observed via its thermal X-ray emission. Galaxy clusters evolve hierarchically through successive mergers and accretion of smaller galaxy groups and clusters. These merger events heat and compress the ICM, while simultaneously generating turbulence and amplifying magnetic fields. 

\subsection{Driving of ICM Turbulence}

A variety of energetic processes drive turbulence in the ICM. The primary driving source is cluster mergers, where collisions of subclusters inject large-scale kinetic energy extending over several hundred kpc. In addition, minor mergers and the motions of member galaxies stir the ICM, generating shear flows and vorticity. Furthermore, outflows from active galactic nuclei (AGN) inflate cavities and drive shocks, thereby supplying turbulence to the surrounding medium \citep[e.g.,][]{2009ApJ...705.1129L, 2017MNRAS.464..210V, 2015ApJ...800...60M}.

Cosmological simulations indicate that turbulent motions contribute a non-negligible fraction to the total energy budget of galaxy clusters, with the ratio of turbulent to thermal energy $E_{\mathrm{turb}} / E_{\mathrm{therm}}$ typically ranging between 5\% and 30\% \citep[e.g.,][]{2009ApJ...705L..90C, 2011MNRAS.418..960V}. During major mergers, turbulent velocities can reach $\sim300~\mathrm{km\,s^{-1}}$ and gradually decay over a few gigayears \citep{2017MNRAS.464..210V}. Thus, the ICM plasma is driven intermittently.

These processes play a central role in the evolution of the ICM. Turbulent motions promote the redistribution of metals, suppress the runaway development of cooling flows, mediate the transport of heat and gas, and control the properties of cosmic rays. Therefore, quantitatively understanding how various energy sources couple to the ICM and how efficiently turbulence is driven across different spatial scales is essential for interpreting the observed thermal and non-thermal phenomena in galaxy clusters.

\subsection{Cascade and Power Spectrum of Turbulence}

\begin{figure}[tbp]
    \centering
	\includegraphics[width=0.7\linewidth]{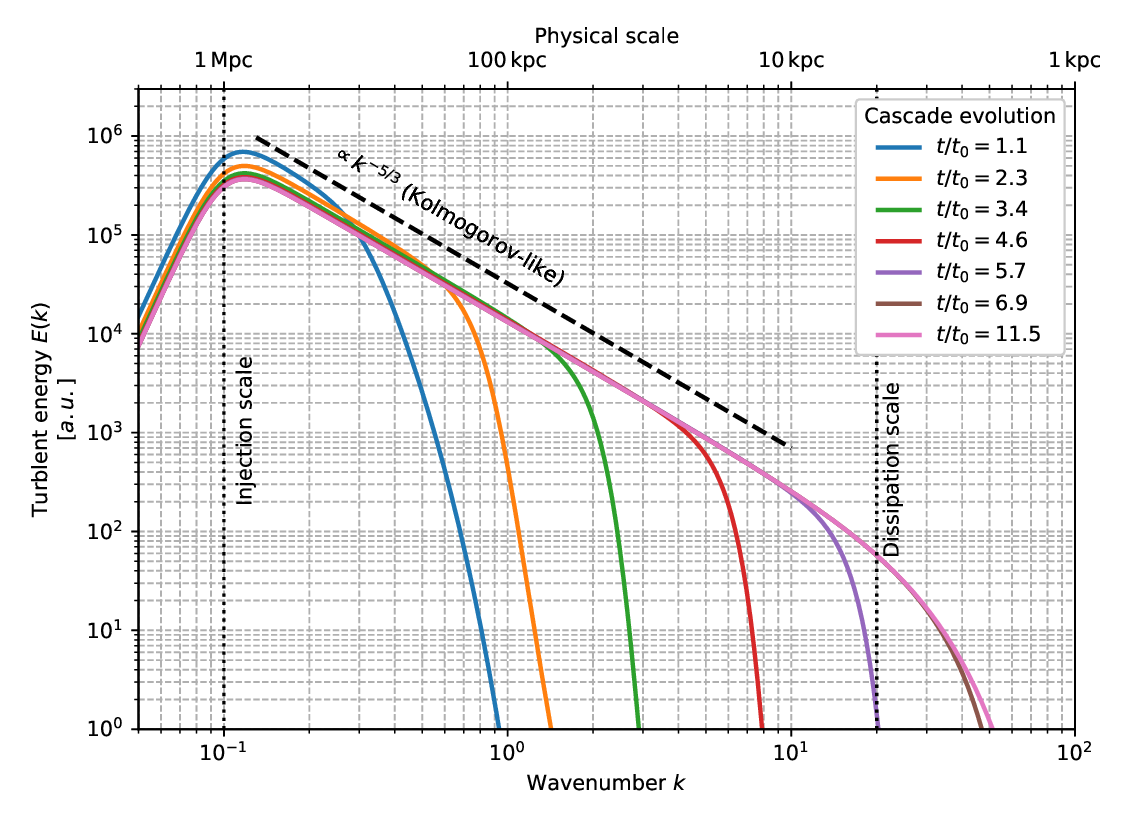}
    \caption{\rev{Conceptual illustration of the time evolution of the turbulent velocity power spectrum driven by a galaxy cluster merger. The elapsed time is normalised by the eddy turnover time at the injection scale, $t_0$. Turbulent energy is injected at large scales and cascades through an inertial range before reaching the dissipation scale, which marks the onset of the turnover of the spectrum. At smaller scales, the power decreases rapidly due to viscous and kinetic dissipation processes. The location of the dissipation scale shown here is illustrative; together with the injection scale, it implies an assumed effective Reynolds number.
}}
    \label{fig:power_spectrum}
\end{figure}

Once injected, the energy of turbulence cascades from large (sub-Mpc) to small (kpc) scales \citep{2004A&A...426..387S,2007mhet.book...85S,2022ApJ...926..183C}. This energy transfer ultimately dissipates at small scales where viscosity and kinetic effects become dominant. Figure~\ref{fig:power_spectrum} schematically illustrates the evolution of the velocity power spectrum driven by the turbulent cascade. In general, the velocity power spectrum can be described as $P(k) \propto k^{-\alpha},$ where $\alpha$ characterises the efficiency of energy transfer across spatial scales. In the simplest hydrodynamic case of incompressible turbulence, $\alpha = 5/3$, corresponding to the Kolmogorov spectrum and a scale-independent energy transfer rate per unit mass. However, ICM is a weakly magnetised plasma, and magnetic fields, compressibility, and plasma microphysical processes may modify the spectral slope from the Kolmogorov expectation. Measuring $\alpha$ therefore provides an important diagnostic of the turbulent cascade and the underlying plasma physics in galaxy clusters.

In the regime of magnetohydrodynamics (MHD), turbulence becomes anisotropic with respect to the local magnetic field direction \citep{1995ApJ...438..763G}. Numerical simulations of cluster turbulence generally show spectra broadly consistent with Kolmogorov-like scaling \citep[e.g.,][]{2018MNRAS.474.1672V, 2015ApJ...810...93P}, although deviations associated with MHD effects and plasma conditions may occur. Observationally, analyses of X-ray surface brightness fluctuations suggest the presence of a near-Kolmogorov cascade \citep{2014Natur.515...85Z, 2018ApJ...854..167G}, while Faraday rotation measurements from radio polarisation observations also indicate magnetic field fluctuations with similar power-law behaviour \citep{2010A&A...513A..30B}.

\subsection{Turbulent Dynamo}

Turbulence in the ICM not only redistributes energy but also amplifies and maintains magnetic fields through the small-scale dynamo mechanism \citep{2004PhRvE..70c6408H,2004ApJ...612..276S,2006MNRAS.366.1437S,2020PhRvF...5d3702S}. Even a weak primordial or seed magnetic field \citep{2016RPPh...79g6901S,2020MNRAS.499.2076S} can be strengthened by the stretching and folding of magnetic field lines driven by turbulent motions \citep{2016ApJ...817..127B,2021PhRvF...6j3701S}. Simulations show that this dynamo amplification saturates when the magnetic energy reaches a few per cent of the turbulent kinetic energy \citep{2008Sci...320..909R,2010ApJ...723..476A,2013MNRAS.429.2469B,2020PhRvF...5d3702S,2021PhRvF...6j3701S}, yielding magnetic field strengths consistent with observed microgauss-level, also inferred from Faraday rotation measurements \citep{2010A&A...522A.105G,2017A&A...603A.122G,2025A&A...694A..44O}. This analysis requires knowledge of the thermal electron density, $n_e$, which can be estimated from X-ray observations of the ICM.

The spatial complexity of these magnetic fields directly affects thermal conduction, gas transport, and the polarisation properties of synchrotron emission. Coherence scales and anisotropies of the magnetic field are manifested in observed depolarisation and rotation measure properties, providing valuable clues to probe the underlying turbulence and magnetic fields. Therefore, magnetic field observations also serve as a powerful, albeit indirect, tool to trace turbulence and its interaction with the plasma.

\subsection{Particle Acceleration}

Furthermore, turbulence can accelerate particles via the second-order Fermi mechanism. Diffuse synchrotron emission in the form of radio halos and radio relics probes cosmic-ray electrons and magnetic fields extending over scales of several hundred kiloparsecs. These diffuse radio sources are commonly observed in merging galaxy clusters, with radio halos occurring preferentially in dynamically disturbed systems. Their prevalence suggests that a significant fraction of the enormous energy released during cluster mergers is converted into non-thermal components. Merger-driven shocks and turbulence are thought to play a central role in this energy conversion, generating and reaccelerating relativistic particles while amplifying magnetic fields throughout the intracluster medium.

The radiative cooling time of cosmic-ray electrons in the ICM is relatively short, $\sim 0.1$~Gyr, and their diffusion in galaxy clusters is limited by magnetic fields (only $\sim 10$--$100$~pc), making large-scale transport of relativistic electrons highly inefficient. Therefore, they lose energy before reaching Mpc scales, which supports the scenario in which turbulence throughout the cluster in situ (re-)accelerates them. Another possibility is that secondary electrons, produced through hadronic interactions of long-lived cosmic-ray protons with thermal ICM protons, generate the observed synchrotron emission. Because the collision frequency is low (collision time $\sim 10$~Gyr), cosmic-ray protons can survive for much longer than electrons and produce secondary electrons throughout the cluster volume, naturally leading to Mpc-scale radio emission. However, this hadronic scenario alone does not explain why giant radio halos are predominantly observed in merging clusters.

\rev{
\subsection{Chapter Outline}
To understand these physical processes, it is essential not only to pursue radio studies with SKA but also to integrate observations from other facilities with complementary capabilities. In this chapter, we focus on the synergy between the SKA and X-ray observations, particularly those enabled by XRISM, in studies of galaxy clusters, and outline a systematic framework toward understanding these underlying physical processes. The remainder of this chapter is organised as follows. In Section 2, we review the X-ray observations of the thermal ICM, focusing on the latest results from XRISM that provide detailed views of merger geometry and the energy spectra of gas turbulence. Section 3 describes the current status of radio observations of non-thermal components, highlighting the discovery of new diffuse structures revealed by SKA pathfinders and the characterisation of magnetic field spectra via rotation measure analysis. In Section 4, we discuss the quantitative synergy between X-ray and radio observations; specifically, we propose practical frameworks to link the properties of magnetic turbulence to synchrotron emission, estimate turbulent scales through time and length scale analyses, and break degeneracies in determining magnetic field and cosmic-ray properties. Section 5 evaluates the observational feasibility and expected performance of the SKA-Mid (AA4) in unravelling these physical processes. Finally, we provide a summary of the chapter in Section 6.
}

\section{X-ray Views of Galaxy Clusters}

X-ray observations reveal the density and temperature structure of the ICM by detecting thermal bremsstrahlung and emission lines from heavy elements. Moreover, precise measurements of the line broadening and shifts of these lines enable direct determination of the line-of-sight gas velocities and their dispersion, providing information on bulk motions and turbulent velocity. X-ray images also clearly display structures such as shock fronts and cold fronts, which can act as potential drivers of turbulence.

In recent years, high-energy-resolution X-ray spectroscopy with instruments such as \textit{Hitomi} and \textit{XRISM} (X-Ray Imaging and Spectroscopy Mission) has enabled direct measurements of turbulence in the ICM. \textit{Hitomi} reported a turbulent velocity dispersion of $\approx$ 164 km s$^{-1}$ in the Perseus cluster \citep{2016Natur.535..117H}, while more recent \textit{XRISM} observations of some cluster measured a turbulent width of $\approx$ 200 km s$^{-1}$ \citep{2025ApJ...993L..11X}. These values appear insufficient to sustain the level of turbulent acceleration required to maintain the observed radio structures \citep{1999LNP...530..106N, 2007ApJ...670L...5B, 2017ApJ...843L..29E}. However, it has been pointed out that such X-ray measurements are weighted by the emissivity of the observed lines, which may bias the inferred average turbulent velocity \citep{2026A&A...705A.129V}. \sout{Also,} for $10^7$–$10^8\,{\rm K}$ ICM plasma, these values correspond to subsonic turbulence.

Early results from XRISM/Resolve have directly measured turbulent velocity dispersions of $100$–$300~\mathrm{km\,s^{-1}}$ in some clusters, such as Perseus, indicating that turbulence accounts for a few per cent of the total energy budget \citep{2025ApJ...993L..11X}. These measurements open the way for a direct X-ray-based investigation of the velocity structure of ICM. 

\rev{
In this section, we briefly review several recent results from XRISM that are relevant to future collaborative studies with the SKA. In particular, XRISM measurements of bulk and turbulent gas motions have revealed the dynamical structure and merger geometry of galaxy clusters in unprecedented detail \citep[e.g.,][]{2025Natur.638..365X,2026ApJ...998..210X}. Such X-ray observations provide crucial guidance on which regions of galaxy clusters should be targeted when deriving the magnetic field configuration and strength distributions with the SKA. Furthermore, the turbulent velocity field inferred from X-ray observations can be directly compared with magnetic field power spectra to be determined by future SKA observations, thereby offering key insights into the evolutionary history of galaxy clusters and the plasma physical processes operating within them.
}


\rev{
\subsection{Detailed Picture of the Cluster Merger Geometry}
Traditional X-ray imaging observations have identified merger-related structures such as shock fronts and cold fronts through surface-brightness and temperature distributions; however, it has been challenging to directly probe the three-dimensional gas motions. In particular, head-on mergers and off-axis (offset) mergers can generate markedly different bulk motions, rotational flows, and turbulence in the post-merger phase. Nevertheless, depending on the projection angle, these distinct merger scenarios can appear remarkably similar in X-ray surface-brightness and temperature maps alone, making it impossible to resolve this geometrical degeneracy without line-of-sight velocity information. The high-resolution spectrometer Resolve onboard XRISM represents a major advance in overcoming these limitations.
}

\rev{
For example, XRISM observations of the galaxy cluster Abell 3667 have, for the first time, clearly revealed systematic variations in the line-of-sight gas velocity along a prototypical cold front, quantitatively demonstrating the presence of large-scale gas flows associated with the merger process \citep{2026ApJ...996L..15O}. These results challenge the previously prevailing view that this cluster is predominantly a head-on merger, and instead indicate signatures of localised turbulence enhancement driven by rotational motions. By resolving such spatially localised variations in both large-scale gas motions and small-scale turbulent motions, XRISM provides the capability to disentangle which physical processes dominate in different regions, even within a single galaxy cluster.
}

\rev{
Similarly, XRISM observations of the core region of the Centaurus galaxy cluster have detected an ordered velocity structure that is likely associated with gas sloshing motions, demonstrating that signatures of past cluster mergers or interactions are preserved in the present-day ICM kinematics \citep{2025Natur.638..365X}. 
Such direct measurements of the velocity field provide powerful constraints on the merger geometry, including the direction of the interaction, impact parameters, and asymmetries in the underlying gravitational potential. In addition, spatially resolved measurements of the turbulent velocity dispersion reveal how merger-driven energy is distributed throughout the ICM and how the level of turbulence varies across the cluster.
}

\rev{
XRISM observations of M87, located at the center of the Virgo cluster, provide a detailed view of the ICM velocity structure in a system where ICM and AGN feedback coexist. High-resolution X-ray spectroscopy has revealed a bulk velocity structure with spatially smooth variations, together with a spatial distribution of turbulence intensity induced by the central AGN activity \citep{2026ApJ...998..210X}. These results demonstrate that AGN feedback plays a key role in shaping the velocity structure of the ICM. Combined with future SKA measurements of magnetic-field structures, such observations will help clarify how AGN-driven gas motions and turbulence amplify, transport, and redistribute magnetic fields within cluster cores.
}

\subsection{Energy Spectra of ICM Velocity}
\rev{
In high-$\beta$ and weakly collisional cluster plasmas, density fluctuations and velocity fluctuations are coupled through the turbulent cascade, allowing the velocity power spectrum to be inferred from the statistical properties of observed X-ray surface-brightness fluctuations \citep{2014A&A...569A..67G}. This method has been applied to multiple galaxy clusters, and the inferred velocity power spectra are broadly consistent with Kolmogorov-like turbulence \citep{2012MNRAS.421.1123C, 2025MNRAS.537.2198L}. However, significant uncertainties remain due to projection effects, assumptions about the thermodynamic state of the ICM, and the modelling and subtraction of large-scale surface-brightness structures, as well as limitations in spatial resolution, preventing definitive conclusions on the detailed spectral shape and dissipation scale.
}

\rev{ 
Recent observations with XRISM suggest that the turbulent velocity spectrum in the Coma cluster may not be fully consistent with a Kolmogorov-like form \citep{2025ApJ...985L..20X}. Instead, the analysis assumes a three-dimensional power spectrum of the form
\begin{equation}
P(k)=P_0
\left[1+(k\ell_{\rm inj})^2\right]^{\alpha/2}
\exp\left[-(k\ell_{\rm dis})^2\right],
\end{equation}
where $\ell_{\rm inj}$ and $\ell_{\rm dis}$ denote the characteristic injection and dissipation scales, respectively. The three-dimensional spectrum is projected into a two-dimensional power spectrum,
\begin{equation}
P_{\rm 2D}(k)
=
\int
P\!\left(\sqrt{k^2+k_z^2}\right)
P_\epsilon(k_z)\, dk_z,
\end{equation}
where $P_\epsilon(k_z)$ is the one-dimensional power spectrum of the normalised emission measure along the line of sight. Using this projected spectrum, the observable velocity structure function (VSF) can be evaluated as
\begin{equation}
{\rm VSF}(r)
\equiv
\left\langle
\left|
\mu_z(\chi+r)-\mu_z(\chi)
\right|^2
\right\rangle
=
4\pi
\int_0^\infty
\left[1-J_0(2\pi kr)\right]
P_{\rm 2D}(k)\, k\, dk.
\label{eq:vsf}
\end{equation}
where $\mu_z$ is the emission-weighted line-of-sight velocity centroid. The observed velocity dispersion is similarly expressed as
\begin{equation}
\sigma_z^2
=
\int
P(k)\left[1-P_\epsilon(k_z)\right] d^3k.
\label{eq:sigma}
\end{equation}
where $\sigma_z$ denotes the emission-weighted line-of-sight velocity dispersion inferred from spectral line broadening. By jointly modeling the VSF and the velocity dispersion, the underlying turbulent power spectrum can be constrained \citep{2025ApJ...985L..20X}.
}

\begin{figure}[tbp]
    \centering
	\includegraphics[width=0.5\linewidth]{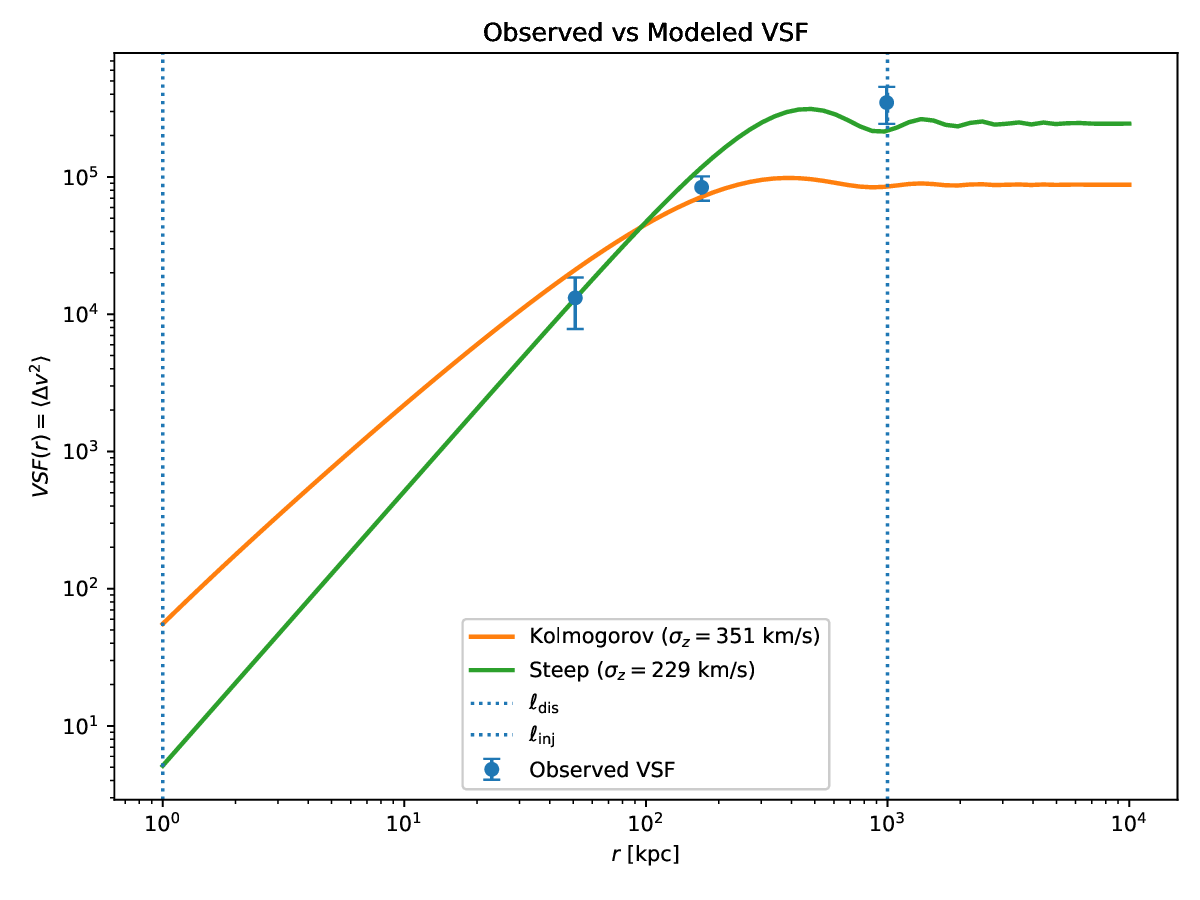}
    \caption{\rev{
    Observed velocity structure function (VSF) of the Coma cluster (blue points) compared with model predictions derived from projected three-dimensional turbulent power spectra. The orange and green curves show models with spectral slopes of $\alpha=-11/3$ (Kolmogorov-like) and $\alpha=-8$, respectively, assuming $\ell_{\rm inj}=1~{\rm Mpc}$ and $\ell_{\rm dis}=1~{\rm kpc}$. 
    }}
    \label{fig:coma_vel_spec}
\end{figure}

\rev{
For example, Figure~\ref{fig:coma_vel_spec} shows the fitting results for the VSF assuming $\ell_{\rm inj}=1~{\rm Mpc}$, $\ell_{\rm dis}=1~{\rm kpc}$, and spectral slopes of $\alpha=-11/3$ and $-8$. Note that, unlike the approach adopted in previous work \citep{2025ApJ...985L..20X}, these results were not obtained by generating multiple realisations of turbulent velocity fields from the assumed power spectrum and then calculating the corresponding VSFs. Instead, neglecting both cosmic variance and systematic uncertainties, we directly computed mock VSFs from the assumed three-dimensional power spectrum using Equations~(1)--(3). For simplicity, the line-of-sight emissivity weighting was approximated by a Gaussian window function, $P_\epsilon(k_z)=\exp[-(k_zL_z)^2]$, with $L_z=1~{\rm Mpc}$. The velocity dispersions calculated from Equation~(4) using the corresponding power spectra between the injection and dissipation scales are $350.5~{\rm km\,s^{-1}}$ for $\alpha=-11/3$ and $228.9~{\rm km\,s^{-1}}$ for $\alpha=-8$. Since the observed velocity dispersion is approximately $200~{\rm km\,s^{-1}}$, the steeper spectrum tends to provide somewhat better agreement with the observations, although the comparison remains qualitative given the simplified treatment adopted here.
}

Nevertheless, this apparent deviation could also arise from observational limitations such as finite spatial resolution and uncertainties associated with X-ray emissivity weighting along the line of sight \citep{2026A&A...705A.129V}, highlighting the need for high-sensitivity, wide-field velocity dispersion mapping. However, the limited field of view of XRISM ($\sim$3'), together with its spatial resolution, hinders direct tracing of the global turbulent velocity field across entire galaxy clusters.

\newpage
\section{Radio Views of Galaxy Clusters}

Radio observations are indispensable for understanding non-thermal structures, and in particular, the importance of low-frequency (< 1 GHz) observations has recently been emphasized. The higher energy of cosmic ray electrons is emitted at higher frequencies but has shorter lifetimes, whereas older electrons cool to lower energies and can only be observed at low frequencies. Therefore, low-frequency observations are especially effective in tracing the remnants of past activity and weak acceleration processes.

Low-frequency radio observations are essential not only for detecting diffuse synchrotron emission but also for providing quantitative constraints on the energy distribution and acceleration efficiency of CR electrons. The synchrotron spectrum reflects the underlying electron energy distribution directly; by examining both the overall spectral index and its curvature (i.e., the energy dependence of the spectral slope), one can infer the acceleration and cooling history of the electron population.

The lifetime of synchrotron-emitting electrons is typically expressed as \citep{2002ARA&A..40..319C, 2014IJMPD..2330007B}:
\begin{equation}
t_{\rm life~time} = 3.2 \times 10^{10} 
\frac{B^{1/2}}{B^2 + B_{\rm CMB}^2} 
\left[(1+z)\,\nu_{\rm obs}\right]^{-1/2}~[{\rm yr}],
\end{equation}
where $B$ is the intracluster magnetic field strength [$\mu$G], $B_{\rm CMB}$ is the equivalent magnetic field corresponding to the cosmic microwave background (CMB) [$\mu$G], $z$ is the cluster redshift, and $\nu_{\rm obs}$ is the target frequency [Hz]. The equivalent CMB magnetic field depends on redshift and is given by
\begin{equation}
B_{\rm CMB} = 3.25 \,(1+z)^2~[\mu{\rm G}].
\end{equation}

Assuming typical cluster conditions of $B \sim 1~\mu$G and $z \sim 0.1$, and substituting an observing frequency of $\nu_{\rm obs} = 100$ MHz, the lifetime of low-frequency-emitting electrons is estimated as
\begin{equation}
t_{\rm life~time} \sim 2 \times 10^8~[{\rm yr}].
\end{equation}
In contrast, for $\nu_{\rm obs} = 1$ GHz, the lifetime is approximately $6 \times 10^7$ yr. Compared to the typical cluster age ($\sim$Gyr), this indicates that high-frequency synchrotron-emitting electrons cool on relatively short timescales. Consequently, the synchrotron spectrum is expected to exhibit a spectral break below $\sim$1 GHz, highlighting the importance of low-frequency observations for tracing the imprints of past acceleration and turbulent processes.

Another major advantage of radio observations is their ability to directly probe magnetic field properties through polarisation and Faraday rotation measurements (Faraday Rotation Measure; RM). Previous RM studies in clusters have primarily relied on measurements towards background radio sources (e.g., \citealt{2001ApJ...547L.111C, 2010A&A...522A.105G}). These observations revealed that $\mu$G-level magnetic fields exist in clusters and provided insights into their statistical properties and spatial scales. Recently, it has become possible to discuss radial profiles of RM across clusters, revealing spatial variations in the magnetic field and providing new insights (e.g., \citealt{2022MNRAS.512.1450R,2024A&A...691A.132R}). 

In recent years, analyses based on Faraday tomography, or the Faraday Depth Function (FDF), have attracted increasing attention \citep{2005A&A...441.1217B}. In this approach, the distribution of polarised intensity as a function of Faraday depth $\phi$ is reconstructed by transforming the observed polarised signal $P(\lambda^2)$ as
\begin{equation}
F(\phi) = \int_{-\infty}^{\infty} P(\lambda^2) \, e^{-2i \phi \lambda^2} \, d\lambda^2.
\end{equation}
The resolution in Faraday depth space, $\delta \phi$, is given by \citep{2005A&A...441.1217B, 2023PASJ...75S..50T}:
\begin{equation}
\delta \phi \approx \frac{2 \sqrt{3}}{\Delta \lambda^2},
\end{equation}
where $\Delta \lambda^2 = \lambda_{\rm max}^2 - \lambda_{\rm min}^2$ represents the wavelength coverage of the observation. Since FDF analysis samples data in $\lambda^2$ space, longer wavelengths (lower frequencies) correspond to widely separated sampling points even for small wavelength differences. This allows one to efficiently detect subtle variations in Faraday depth and extended magnetic structures, providing a powerful means to probe turbulence and magnetic field scales within clusters. Therefore, low-frequency and wideband radio telescopes such as SKA are particularly well suited for such studies.

\rev{
In the following subsections, we briefly review recent progress in radio observations of galaxy clusters, focusing in particular on newly revealed radio structures enabled by high-sensitivity observations with SKA precursor facilities.
}

\subsection{Discovery of New Radio Structures}
\begin{figure}[tbp]
    \centering
	\includegraphics[width=\linewidth]{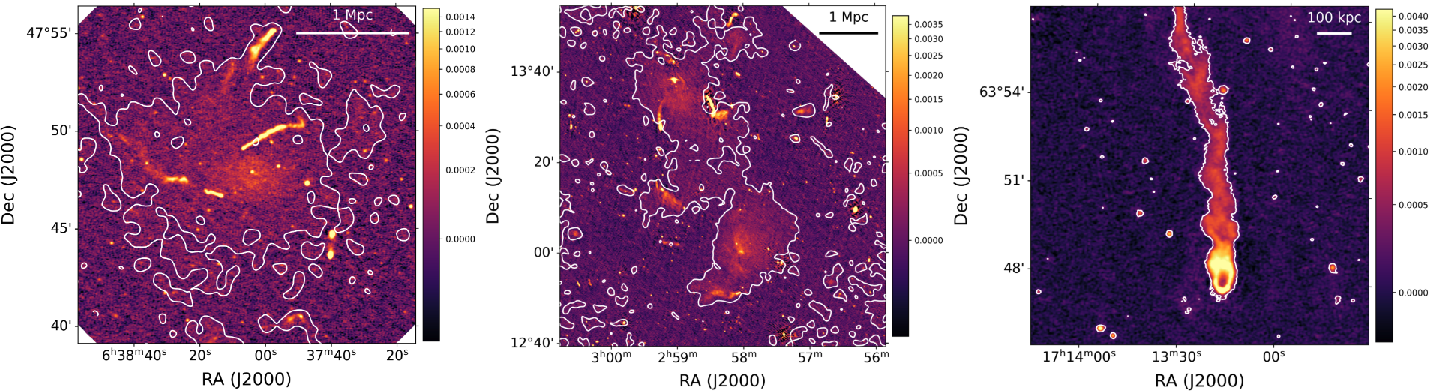}
    \caption{\rev{
    Various newly discovered radio structures identified with SKA pathfinder observations. All color images show the high-resolution LOFAR LoTSS DR3 data, while the contours represent smoothed radio intensity distributions after subtraction of compact sources in the corresponding fields. Scale bars are shown in the upper-right corners of the panels. (Left) A megahalo in ZwCl~0634.1+4750. (Center) A radio bridge connecting Abell~399 and Abell~401. (Right) A head--tail radio galaxy in Abell~2255. 
    }}
    \label{fig:radio_image_from_SKApathfinders}
\end{figure}

Wideband, high-sensitivity low-frequency radio observations with SKA precursor facilities such as LOFAR, MeerKAT, and ASKAP have started to reveal faint synchrotron emission and magnetic field structures that were previously inaccessible. As a result, a more detailed picture of the non-thermal universe within galaxy clusters is emerging. Figure~\ref{fig:radio_image_from_SKApathfinders} shows examples of these radio images.

\subsubsection{Mega-halos}
Recent low-frequency radio observations have revealed extremely extended, faint radio structures, so-called “mega-halos,” in the outskirts of several galaxy clusters \citep{2022Natur.609..911C}. These radio components are raised by turbulence, suggesting that turbulent energy in cluster outskirts may be sufficient to drive particle acceleration. However, the spatial extent over which such large-scale turbulence is maintained, and whether its strength is sufficient to excite radio emission, remains observationally uncertain.

Currently, in the outskirts of galaxy clusters, direct measurements of turbulent velocities with XRISM are challenging due to sensitivity limitations. In the future, long-exposure observations enabling measurements of gas velocity dispersion in these regions with XRISM would allow a direct comparison with radio mega-halo areas, representing a significant step toward a unified understanding of non-thermal energy injection processes.

\subsubsection{Radio Bridges in Early-stage Merging Cluster}
Recent studies have reported faint radio bridge structures in early-stage merging clusters, corresponding to phases preceding the formation of major radio halos or relics \citep[e.g.,][]{2019Sci...364..981G, 2019NatAs...3..838G, 2023PASJ...75S.138K}. These structures are interpreted as emission from electrons accelerated by turbulence induced after shock passage, offering observational insight into the initial stages of turbulence development.

However, bridge emission is extremely faint, and separation from radio galaxies and background emission is challenging. As a result, statistical samples and confirmation of their origin remain limited. Future wideband, high-sensitivity observations with SKA will be crucial to accurately characterise the spatial and spectral properties of these faint emissions, providing key insights into the generation and decay of turbulence in merging galaxy clusters.

\rev{
\subsubsection{Head-tail galaxies}
Head-tail radio galaxies were identified in the late 1960s through early radio surveys \citep[e.g.,][]{1968MNRAS.138....1R}, and were subsequently recognised as cluster-associated sources shaped by ram pressure from the ICM. Such morphologies directly reflect the relative motion between galaxies and the ICM, making head–tail radio galaxies sensitive probes of the local dynamical environment within galaxy clusters. 
In this sense, the morphology of head–tail galaxies provides a valuable opportunity to investigate how magnetic fields are stretched, amplified, and mixed by shear flows and turbulence in cluster environments. Future comparisons with XRISM measurements of the ICM velocity field may provide a direct link between the observed radio morphology and the underlying gas dynamics responsible for shaping these magnetic structures.
}

\rev{
For example, the galaxy cluster Abell 3322 hosts a giant, grand-design head–tail radio galaxy near its center. Exploiting the high angular resolution of SKA precursor observations, a peculiar ``Omega ($\Omega$) structure'' has been discovered and its origin has been investigated in detail \citep{2026PASJ...78..137K}.
Multi-frequency radio observations and spectral analyses suggest that this structure arises from a combination of Doppler boosting of the jets, radiative aging of cosmic-ray electrons, and the formation of twin vortices in a low Reynolds number flow within the ICM. These results demonstrate that head-tail galaxies provide an important laboratory for disentangling the interplay between turbulence, shock waves, and magnetic field structures in cluster environments.
}


\rev{
\subsection{Polarimetric Probes of Intracluster Magnetic Fields}
}
High-sensitivity polarisation observations with the SKA provide detailed information on the spatial distribution and anisotropy of intracluster magnetic fields, complementing velocity measurements obtained from X-ray observations. In particular, high-density rotation measure (RM) sampling and improved Faraday depth resolution enable direct tests of theoretical predictions from numerical MHD simulations, such as anisotropic magnetic field amplification and Alfvén-wave–driven turbulent dissipation.

Previous studies based on Faraday rotation measurements have shown that the magnetic field power spectrum in galaxy clusters is broadly consistent with a Kolmogorov-like form \citep{2010A&A...513A..30B}. This result supports a scenario in which turbulence and magnetic fields in the ICM co-evolve through an MHD turbulent cascade. Furthermore, observations suggest that the magnetic field correlation length is comparable to the characteristic turbulence scale, typically of order 10–100 kpc \citep{2017A&A...603A.122G}. 

\rev{
Recent results from SKA precursor surveys have demonstrated that dense RM grids can statistically detect magnetised thermal gas associated with galaxy groups and their surrounding large-scale environments, extending to several times the splashback radius \citep{2024MNRAS.533.4068A}. These studies highlight the potential of RM-based approaches to probe magnetic fields beyond cluster cores, complementing X-ray and SZ observations.
}

\rev{
To reconstruct three-dimensional cluster-scale magnetic field structures, dense sampling of the RM grid is essential. This requires increasing the surface density of background polarised sources to several hundred per square degree, a level that is already being approached with current precursor facilities such as MeerKAT, and is expected to be significantly enhanced with SKA-class instruments. In addition, resolving magnetic structures along the line of sight demands Faraday depth resolutions reaching sub-rad m$^{-2}$ levels in wide-band low-frequency observations. Consequently, the SKA is expected to play a decisive role in the tomographic reconstruction of magnetic field energy spectra and anisotropies in galaxy clusters.
}

\section{Synergy between X-ray and Radio Observations}
A comprehensive understanding of ICM turbulence and magnetic field requires a multi-wavelength approach that combines both X-ray and radio observations. X-ray measurements provide a direct probe of the velocity dispersion of the thermal component, whereas radio observations reveal the structure of magnetic fields (assuming information about the thermal electron density is known or assumed) and non-thermal energy components. By combining these two modalities, it becomes possible to quantitatively assess the energy distribution and dissipation processes of turbulence and magnetic fields within the ICM.

Thus, the combined analysis of X-ray and radio data represents the most effective method to observationally probe the coupled evolution of turbulence, magnetic fields, and cosmic rays. The synergy between SKA and XRISM is expected to bring a new paradigm in the study of ICM physics. In addition, emerging observational evidence from several galaxy clusters suggests a possible correlation between the measured level of turbulence from X-ray observations, including recent XRISM measurements, and the observed radio emission intensity (Omiya, 2026, PhD thesis). This trend, while still based on a limited sample, indicates that regions with enhanced turbulent activity tend to exhibit stronger non-thermal radio emission, providing further support for a tight connection between turbulence and magnetic field amplification in the ICM.

\rev{
In the following subsections, we introduce several key topics that can be explored by jointly analysing X-ray and radio observations.
}

\rev{
\subsection{Breaking Degeneracies: Magnetic Field Strength and Cosmic Ray Energetics}
RM depends on the product of the electron density, $n_e$, and the line-of-sight magnetic field component, $B_\parallel$, making it difficult to separate the contributions of these two quantities using radio observations alone. However, the gas distribution in galaxy clusters is well described by a $\beta$-model, and once the spatial distribution of $n_e$ is determined through X-ray observations, the combination with high-density RM grids provided by the SKA enables a more accurate extraction of the spatial strength and structure of the intracluster magnetic field.
}

\rev{
Furthermore, by applying such independently constrained magnetic field strengths to the analysis of synchrotron emission, it becomes possible to break the degeneracy between the magnetic field strength and the number density of relativistic electrons, $n_{\rm CRe}$, that is inherently embedded in the radio intensity, and to quantitatively evaluate the energy distribution of cosmic-ray electrons. In addition to magnetic field estimates inferred from synchrotron intensity, upper limits on non-thermal hard X-ray emission obtained from X-ray observations provide independent constraints on the magnetic field strength through inverse Compton scattering. In this way, constraining the same physical quantities using multiple, complementary observational techniques and cross-validating them against each other is critically important for reducing systematic uncertainties and for achieving a robust understanding of the physics of non-thermal components in galaxy clusters.
}

\rev{
\subsection{Turbulence Intensity as a Driver of Synchrotron Emission} 
The synchrotron brightness intrinsically depends on both the magnetic field strength and the energy distribution of relativistic electrons. For this reason, it is fundamentally difficult to uniquely determine the particle re-acceleration efficiency or the magnetic field energy density from radio observations alone. XRISM enables spatially resolved measurements of the turbulent velocity dispersion in the ICM. This capability provides a key observational test of the turbulent re-acceleration model, which has been proposed as the origin of diffuse radio emission such as radio halos. 
}


\revtwo{
We can derive the electron acceleration and magnetic field amplification efficiencies from the turbulent energy budget as follows \citep{2022SciA....8.7623B}.
The turbulent energy density is
\begin{equation}
U_{\rm turb} = \frac{1}{2}\rho \sigma_v^2,
\label{eq:Uturb}
\end{equation}
where $\rho$ is the gas mass density and $\sigma_v$ is the turbulent velocity dispersion measured by XRISM.
Assuming a characteristic turbulence scale $L$, the eddy turnover time is
\begin{equation}
t_{\rm edd} = \frac{L}{\sigma_v}.
\label{eq:tedd}
\end{equation}
The turbulent energy dissipation rate per unit volume is therefore
\begin{equation}
\dot{E}_{\rm turb}
=
\frac{U_{\rm turb}}{t_{\rm edd}}
=
\frac{1}{2}\rho \frac{\sigma_v^3}{L}.
\label{eq:Eturb}
\end{equation}
The magnetic-field amplification efficiency is defined as the ratio of the magnetic energy density to the turbulent energy density,
\begin{equation}
\eta_B \equiv \frac{U_B}{U_{\rm turb}},
\end{equation}
where
\begin{equation}
U_B = \frac{B^2}{8\pi}.
\end{equation}
Substituting Eq.~(\ref{eq:Uturb}) gives
\begin{equation}
\eta_B
=
\frac{B^2/(8\pi)}
     {(1/2)\rho\sigma_v^2}
=
\frac{B^2}{4\pi \rho \sigma_v^2}.
\label{eq:etaB}
\end{equation}
Thus, $\eta_B$ is determined directly from the magnetic-field strength inferred from RM observations together with the gas density and turbulent velocity measured by X-ray observations.
The electron acceleration efficiency is defined as
\begin{equation}
\eta_{\rm acc}
\equiv
\frac{\dot{E}_{\rm NT}}
     {\dot{E}_{\rm turb}},
\label{eq:etaacc_def}
\end{equation}
where $\dot{E}_{\rm NT}$ denotes the non-thermal energy dissipation rate. Following Botteon et al.~(2022), the non-thermal power can be estimated from the observed synchrotron luminosity $L(\nu)$ as
\begin{equation}
\dot{E}_{\rm NT}
\simeq
\frac{\xi L(\nu)}{V}
\left(
1+\frac{B_{\rm CMB}^2}{B^2}
\right),
\label{eq:ENT}
\end{equation}
where $\xi$ converts the monochromatic luminosity into the total synchrotron luminosity and $V$ is the emitting volume.
Combining Eqs.~(\ref{eq:Eturb})--(\ref{eq:ENT}) yields
\begin{equation}
\eta_{\rm acc}
\simeq
\frac{\xi L(\nu)}
     {V \left(\frac{1}{2}\rho\sigma_v^3/ L \right)}
\left(
1+\frac{B_{\rm CMB}^2}{B^2}
\right).
\label{eq:etaacc}
\end{equation}
Equation~(\ref{eq:etaacc}) clearly shows that $\eta_{\rm acc}$ depends not only on the observed radio luminosity, gas density, and turbulent velocity dispersion, but also linearly on the assumed turbulence scale $L$.
}


\revtwo{
Therefore, once the characteristic turbulence scale $L$ and the emitting volume $V$ are assumed, all quantities entering Eqs.~(\ref{eq:etaB}) and (\ref{eq:etaacc}) can be constrained observationally. Specifically, $\rho$ and $\sigma_v$ are measured from X-ray observations, while $B$ and $L(\nu)$ are obtained from radio observations. This enables direct estimates of both $\eta_B$ and $\eta_{\rm acc}$ from joint X-ray and radio data sets. Such an approach is currently being tested for Abell~1060 using XRISM and MeerKAT data (Kurahara et al., in prep.).
}

\begin{figure}[tbp]
    \centering
	\includegraphics[width=0.75\linewidth]{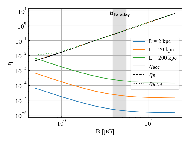}
    \caption{
    Estimated electron acceleration efficiency, $\eta_{\rm acc}$, and magnetic field amplification efficiency, $\eta_B$, as functions of the assumed magnetic field strength in the Coma cluster. The shaded region indicates the $1\sigma$ confidence interval of the central magnetic-field strength estimated by \citet{2010A&A...513A..30B}.
    }
    \label{fig:estimation_of_accel_B_eff}
\end{figure}

\rev{
Figure~\ref{fig:estimation_of_accel_B_eff} shows the estimated values of $\eta_{\rm acc}$ and $\eta_B$ calculated using Eqs.~(\ref{eq:etaacc}) and (\ref{eq:etaB}) as functions of the assumed magnetic-field strength of Coma cluster. The dependence on the adopted turbulence scale is illustrated for $L = 2$, 20, and 200~kpc. As a representative example based on the Coma cluster, we adopt a turbulent velocity dispersion of $\sigma_v = 217~{\rm km~s^{-1}}$ and an electron density of $n_e = 3.4\times10^{-3}~{\rm cm^{-3}}$. We further estimated that the flux density within the region where the velocity dispersion is measured is $0.1~{\rm Jy}$ at 144 MHz, as inferred from the LoTSS data. The magnetic field strength in the Coma cluster has been independently constrained through Faraday rotation measure analyses, yielding $B \simeq 4.7~\mu{\rm G}$ \citep{2010A&A...513A..30B}. Using this value, we obtain $\eta_B \sim 10\%$ and $\eta_{\rm acc} \sim 0.003\%$ for L = 20~kpc case.
}

\rev{
\subsection{Constraining Turbulent Scales through CR Cooling and Acceleration Timescales}
The re-acceleration timescale of turbulent, $t_{\rm acc}$, depends on the turbulent Mach number, $\mathcal{M}$, and the characteristic driving scale, $L$ \citep[e.g.,][]{2016MNRAS.458.2584B}. On the other hand, the cutoff frequency, $\nu_c$, observed at the high-frequency end of the radio synchrotron spectrum suggests that the radiative cooling timescale of cosmic-ray electrons at that energy, $t_{\rm cool}$, becomes comparable to the re-acceleration timescale. Although $t_{\rm cool}$ depends on the magnetic field strength of the system, once the magnetic field strength and cutoff frequency are constrained, a comparison between $t_{\rm cool}$ and $t_{\rm acc}$ can potentially provide constraints on the characteristic turbulence scale responsible for particle acceleration.
}

\begin{figure}[tbp]
    \centering
	\includegraphics[width=0.6\linewidth]{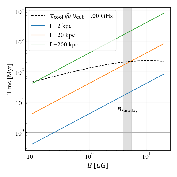}
    \caption{
    Comparison between the radiative cooling timescale of cosmic-ray electrons, $t_{\rm cool}$, and the turbulent acceleration timescale, $t_{\rm acc}$, for the Coma cluster. The shaded region indicates the $1\sigma$ confidence interval of the central magnetic-field strength estimated by \citet{2010A&A...513A..30B}.
    }
    \label{fig:constraining_of_turb_scale}
\end{figure}

\rev{
Figure~\ref{fig:constraining_of_turb_scale} shows an example for the Coma cluster. Since previous studies have suggested that the synchrotron spectrum of the Coma radio halo exhibits a cutoff at frequencies of a few GHz \citep{2003A&A...397...53T}, we assume a cutoff frequency of $1~{\rm GHz}$. Adopting the magnetic field strength inferred from RM analyses, $B=4.7~\mu{\rm G}$, we find that the cooling timescale becomes comparable to the acceleration timescale expected for turbulence with a characteristic scale of $\sim20~{\rm kpc}$, assuming the turbulent velocity measured by XRISM observations of the Coma cluster. These estimates rely on several physical assumptions and simplified treatments; the associated uncertainties should be interpreted with caution.
}

\section{Observational Feasibility with SKA}

\subsection{Detection of Faint Emission}

To diagnose turbulence in galaxy clusters via non-thermal emission, it is essential to detect synchrotron radiation over a broader range of cluster parameters. The SKA AA4 Band 1 is expected to significantly outperform current facilities in both spatial resolution and sensitivity, reaching a sensitivity of $1\sigma \sim 0.5~\mu\mathrm{Jy~beam^{-1}}$ with 10 hours of integration under natural weighting (see SKA Mid Sensitivity Calculator Documentation). This sensitivity represents roughly an order-of-magnitude improvement over current LOFAR and uGMRT observations, enabling direct detection of faint halo emission and peripheral megahalo components that were previously near the detection threshold \citep[e.g.,][]{2020PhRvL.124e1101B, 2022Natur.609..911C}.

In particular, radio emission is expected to be detectable in moderately massive clusters where radio halos have not been observed, as well as in clusters in the early stages of merging. To provide a more quantitative estimate, the flux density per beam, $S_\mathrm{beam}$ (Jy/beam), can be calculated from the total radio power of a halo, $P_\nu$ (W Hz$^{-1}$), as

\begin{equation}
S_\mathrm{beam} = \frac{P_\nu}{4 \pi D_L^2 (1+z)^{\alpha-1}} \cdot \frac{\Omega_\mathrm{beam}}{\Omega_\mathrm{halo}},
\end{equation}

where $D_L$ is the luminosity distance, $\alpha$ is the spectral index (here assumed to be $\alpha = -1.0$), and $\Omega_\mathrm{beam}$ and $\Omega_\mathrm{halo}$ are the solid angles of the beam and halo, respectively. These can be computed as

\begin{equation}
\Omega_\mathrm{beam} = \theta_\mathrm{beam}^2, \quad
\Omega_\mathrm{halo} = \theta_\mathrm{halo,x} \times \theta_\mathrm{halo,y}.
\end{equation}

For example, a halo with a physical size of $500 \times 500~\mathrm{kpc^2}$ at redshift $z=0.1$ corresponds to an angular size of approximately $(5.8~\mathrm{arcmin})^2$ under standard cosmological assumptions. Assuming a beam size of $5 \times 5~\mathrm{arcsec^2}$, the flux density per beam is estimated as shown in Table~\ref{tab:flux_density_example}.

\begin{table}[h]
\centering
\caption{Example of flux density per beam for a halo at redshift $z=0.1$ observed with a $5 \times 5~\mathrm{arcsec^2}$ beam.}
\label{tab:flux_density_example}
\begin{tabular}{cc}
\hline
Total radio power $P_{1.4~\mathrm{GHz}}$ [W/Hz] & Flux per beam $S_\mathrm{beam}$ [µJy/beam] \\
\hline
$1 \times 10^{23}$ & 0.11 \\
$3 \times 10^{23}$ & 0.33 \\
$1 \times 10^{24}$ & 1.10 \\
\hline
\end{tabular}
\end{table}

As these calculations indicate, faint halos with total radio power of $P_{1.4~\mathrm{GHz}} \sim 10^{23}$ W/Hz, which have not been previously detected, are within reach of SKA AA4 standard sensitivity and can be observed with a reasonable integration time.  

For clusters with already known radio halos, an important goal of SKA is to spatially resolve and detect the peripheral megahalo components. Currently detected megahalos have radio powers on the order of $10^{23}$ W/Hz \citep{2022Natur.609..911C}, comparable to classical halos. However, since megahalos are more extended than classical halos, their surface brightness is lower, making detection more challenging; nevertheless, they are likely to be detectable in several massive clusters. In addition to detection, it is crucial to determine radial profiles of the emission, which will allow us to characterise where magnetic fields play a significant role in the synchrotron radiation and where they do not—ideally, magnetic fields contribute across the entire cluster volume.

Moreover, accurate separation of compact components from central galaxies or AGN is required, demanding a dynamic range of at least $10^{5}$. This high dynamic range is necessary because typical central galaxies can have radio flux densities on the order of several Jy, whereas peripheral halo emission is often at the $\mu$Jy/beam level. Simulations indicate that SKA-MID, with its excellent $uv$ coverage and calibration capabilities, can achieve sufficient imaging fidelity to reliably subtract these compact sources. For instance, a 2-hour integration at L-band is expected to detect diffuse halo emission down to $\sim 1~\mu$Jy/beam at 2 arcsec resolution, while adequately mitigating contamination from central sources. Such performance will enable robust characterisation of halo morphology and low surface-brightness features for the first time.

\subsection{Density of RM Grids and Spatial Resolution}

RM grid observations using background polarized sources are one of the most effective methods for statistically analyzing turbulence and magnetic field structures in galaxy clusters. To detect magnetic field fluctuations induced by turbulence on scales of tens of kpc, it is desirable to obtain at least several tens of independent RM measurements per cluster. For a typical galaxy cluster with an apparent diameter of $\sim30'$, this requirement translates to a surface density of background polarized sources of at least $50~\mathrm{deg^{-2}}$.  

Current polarization surveys, such as POSSUM (ASKAP), achieve source densities of $25$–$30~\mathrm{deg^{-2}}$. In contrast, SKA-MID Band 2 is expected to enable high-density RM grids of $100~\mathrm{deg^{-2}}$ with roughly 10-hour integration per field \citep{2020Galax...8...53H}. Such a dense RM sampling will allow statistical analyses of the RM distribution across the entire cluster on kpc scales, for instance using structure functions with higher-order stencils \citep{2023MNRAS.518..919S, 2024MNRAS.533.1875S}, making it feasible for the first time to directly estimate spatial anisotropies and the characteristic scales of magnetic turbulence. Furthermore, wideband polarization observations will achieve Faraday depth resolutions of $\sim 0.4~\mathrm{rad\,m^{-2}}$ in Band 1 and $\sim 4~\mathrm{rad\,m^{-2}}$ in Band 2. These resolutions are sufficient to resolve typical magnetic structures in large galaxy clusters, and enhance the potential to separate the magnetic structures in cluster cores from those in the outskirts.

\rev{
\subsection{From SKA Precursors to the Full SKA}
In addition to the capabilities of the full SKA, important progress is already being achieved through its precursor facilities, such as MeerKAT and ASKAP. Recent works have demonstrated the power of deep, high-sensitivity radio observations in revealing diffuse synchrotron emission and complex magnetic field structures in nearby cluster environments \citep[e.g.,][]{2026arXiv260418338L}. These studies provide early constraints on the spatial distribution of relativistic electrons and magnetic fields, as well as their connection to the dynamical state of the intracluster medium. However, the limited sky coverage, sensitivity, and source density of RM grids with current precursors still restrict the ability to statistically characterise turbulence and magnetic field fluctuations across a wide range of scales. The full SKA will significantly enhance these capabilities by increasing RM source densities by orders of magnitude, improving angular resolution and sensitivity, and enabling systematic, high-fidelity mapping of magnetic field structures in galaxy clusters. This transition from precursor to SKA-era observations will therefore be critical for extending current results into a quantitative, statistically robust framework, particularly when combined with XRISM measurements of ICM velocity fields.
}

\section{Summary}

Turbulence, magnetic fields, and cosmic rays constitute key non-thermal components of the ICM, shaping the thermodynamic and dynamical evolution of galaxy clusters. Recent advances in both X-ray and radio observations have begun to provide direct constraints on the driving, cascade, and dissipation of turbulence, as well as on the amplification and structure of cluster magnetic fields. High-resolution X-ray spectroscopy with \textit{XRISM} has enabled measurements of turbulent velocities and bulk motions, while low-frequency, wide-band radio observations with SKA precursors have revealed diffuse synchrotron structures, Faraday-depth complexity, and large-scale magnetic field fluctuations. These complementary approaches now allow us to probe the coupling between thermal and non-thermal energy in clusters with unprecedented detail.

Despite this progress, several challenges remain. Measurements of turbulent spectra in the ICM are currently limited by instrumental resolution and emissivity-weighting effects, while the magnetic field power spectrum relies on dense RM grids that are only beginning to be realized. The origin of large-scale non-thermal structures—such as mega-halos, cold-front–associated turbulence, and early-stage radio bridges—also remains uncertain due to the extreme faintness of these components and the difficulty of disentangling them from embedded or background sources. Furthermore, existing observations suggest that turbulence inferred from X-rays may be insufficient to account for the full energetics required for particle re-acceleration, highlighting the need for spatially resolved, multi-wavelength diagnostics capable of capturing the full three-dimensional turbulence field.

The forthcoming SKA era, combined with continued \textit{XRISM} operations, will mark a transformative stage in our understanding of cluster turbulence. Wide-field velocity-dispersion mapping, dense RM grids, and high-sensitivity, low-frequency polarimetry will enable direct reconstruction of turbulent and magnetic field properties from cluster cores to outskirts. Together, these observations will provide the first comprehensive view of how turbulence is generated, cascades across scales, and interacts with cosmic rays and magnetic fields. Such progress will be essential for establishing a unified physical framework for the non-thermal ICM and, ultimately, for elucidating the role of turbulence in the broader context of cosmic structure formation.

\bibliographystyle{abbrvnat-maxbibnames4}
\bibliography{SKA_SB_KK2025}
\end{CJK*}
\end{document}